\documentclass[preprint,5p]{elsarticle}
\usepackage{stmaryrd}
\usepackage{amsmath}
\usepackage{amssymb}
\usepackage{tabularx}
\usepackage{subfigure}
\usepackage{color}
\usepackage{indentfirst}
\SetSymbolFont{stmry}{bold}{U}{stmry}{m}{n}
\biboptions{sort&compress}
\usepackage{graphicx}
\usepackage{enumitem}
\numberwithin{equation}{section}

\renewcommand{\vec}[1]{\mathbf{#1}}

\def\subFV{\scriptscriptstyle{FV}}
\def\subFS{\scriptscriptstyle{FS}}
\def\subVS{\scriptscriptstyle{VS}}

\begin{document}

\begin{frontmatter}
\title{Application of Onsager's variational principle to the dynamics \\
of a solid toroidal island on a substrate}

\author[1,2]{Wei Jiang\corref{8}}
\address[1]{School of Mathematics and Statistics, Wuhan University, Wuhan, 430072, China}
\address[2]{Computational Science Hubei Key Laboratory, Wuhan University, Wuhan, 430072, China}
\ead{jiangwei1007@whu.edu.cn}
\cortext[8]{Corresponding author at: School of Mathematics and Statistics,
Wuhan University, Wuhan, 430072, China.}

\author[3]{Quan Zhao}
\address[3]{Department of Mathematics, National University of
Singapore, Singapore, 119076}

\author[4]{Tiezheng Qian}
\address[4]{Department of Mathematics,
Hong Kong University of Science and Technology, Clear Water Bay, Kowloon, Hong Kong}

\author[5,6,7]{David J. Srolovitz}
\address[5]{Departments of Materials Science and Engineering,
University of Pennsylvania, Philadelphia, PA 19104, USA}
\address[6]{Departments of Mechanical Engineering and Applied Mechanics,
University of Pennsylvania, Philadelphia, PA 19104, USA}
\address[7]{Departments of Materials Science and Engineering,
City University of Hong Kong, Kowloon, Hong Kong}

\author[3]{Weizhu Bao}


\begin{abstract}

In this paper, we consider the capillarity-driven evolution of a solid toroidal island on a flat rigid substrate, where mass transport is controlled by surface diffusion. This problem is representative of the geometrical complexity associated with the solid-state dewetting of thin films on substrates.
We apply Onsager's variational principle to develop a general approach for describing surface diffusion-controlled problems. Based on this approach, we derive a simple, reduced-order model and obtain an analytical expression for the rate of island shrinking and validate this prediction by numerical simulations based on a full, sharp-interface model.
We find that the rate of island shrinking is proportional to the material constants $B$ and
the surface energy density $\gamma_0$, and is inversely proportional to the island volume $V_0$.
This approach represents a general tool for modeling interface diffusion-controlled morphology evolution.

\end{abstract}

\begin{keyword}
Solid-state dewetting, Onsager's variational principle, surface diffusion, shrinking instability, moving contact line.
\end{keyword}
\end{frontmatter}

\section{Introduction}

Thin solid films are often metastable or unstable even if they are at a temperature well below their melting points
and can agglomerate or dewet to form small islands on the substrate due to surface tension and capillary effects.
This process, often referred to as solid-state dewetting~\cite{Thompson12,Leroy16}, is ubiquitous in thin film processing and application, and has demonstrated interesting geometric complexities, e.g.,
a thin film producing a series of growing holes \cite{Srolovitz86b,Ye11a,Amram12,Zucker16},
retracting edges~\cite{Wong00,Dornel06,Kim13}, or  breaking up into small particles~\cite{Jiang12,Kim15,Bao17}.

Unlike wetting/dewetting of liquid films, here, the mass transport is usually dominated by surface diffusion rather than fluid dynamics. The governing equation for the kinetic evolution of the film/vapor interface with isotropic surface energy was given by Mullins~\cite{Mullins57},
\begin{equation}
v_n=B\gamma_0\,\Delta_s\mathcal{H},
\label{eqn:mullins}
\end{equation}
where $v_n$ is the normal velocity of the film/vapor interface (surface), $B={D_s \nu\,\Omega_0^2}/{k_BT}$ is a material constant,
$\gamma_0$ is the isotropic surface energy density, $D_s$ is the surface diffusivity, $k_B T$ is the thermal energy,
$\nu$ is the number of diffusing atoms per unit area, $\Delta_s$ is the Laplace-Beltrami operator,
and $\mathcal{H}$ represents the mean curvature of the interface~\cite{Pressley10,Pruss16}.

\begin{figure}[!htp]
\centering
\includegraphics[width=0.48\textwidth]{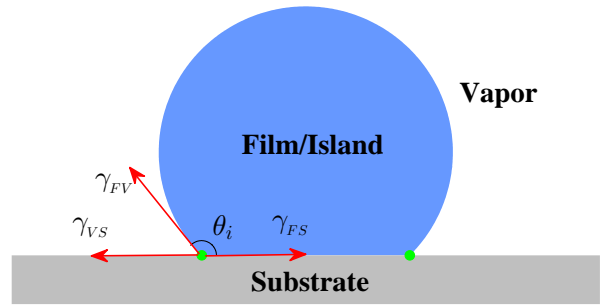}
\caption{The cross-section profile of the equilibrium shape of a solid film/island on a substrate.}
\label{fig:youngequation}
\end{figure}

In addition, solid-state dewetting involves the motion of contact lines.
Recently, morphological evolution in three phase systems with contact lines has attracted significant attention in many different research communities, e.g., Rayleigh instability in the presence of substrates~\cite{Mccallum96,Wong97,Kirill02,Savina04,Gurski06,Roper10}, wetting/dewetting in  batteries~\cite{Chen11,Abdeljawad14}, heterogeneous nucleation at walls~\cite{Granasy07,Warren09}.
As illustrated in Fig.~\ref{fig:youngequation}, the contact line is a triple line where three phases (solid film, vapor, and solid substrate) meet.
The static, equilibrium  contact angle  $\theta_i$ satisfies the well-known Young-Dupree equation~\cite{Young1805}, i.e.,
\begin{equation}
\gamma_{_{\subVS}} = \gamma_{_{\subFS}} + \gamma_{_{\subFV}}\cos\theta_i,
\label{eqn:youngequation}
\end{equation}
where $\gamma_{\subFV}$ (often written as $\gamma_0$), $\gamma_{\subFS}$ and $\gamma_{\subVS}$ are the surface energy densities of the film/vapor, film/substrate and vapor/substrate interfaces, respectively.

Capillary instabilities (e.g., occur in solid-state dewetting) are especially well-known in liquid systems, e.g., thin liquid cylindrical jets as discussed by Plateau~\cite{Plateau73} and Lord Rayleigh~\cite{Rayleigh78}.
They have shown that a small volume-preserving sinusoidal perturbation with wavelength exceeding the circumference of the cylinder can grow exponentially in order to reduce the surface energy, and consequently, the cylinder will break up into a series of small spherical islands. Recently, the Rayleigh instability for more complex geometries (such as liquid toroids on a substrate) have attracted considerable interest in the physics and materials science community where it has received both experimental (e.g.,\cite{Pairam09,Mcgraw10,Wu10}) and theoretical (e.g.,\cite{Wu03,Nguyen12,Mehrabian13}) attention. In analogy to the cylinder, a toroid can also exhibit a Rayleigh instability in the azimuthal direction.
On the other hand, its radial curvature also produces a variation of the mean curvature,
forcing the toroid to shrink towards its own center, leading to its collapse into a compact object and eventually to a section of a sphere~\cite{Yao11, Fragkopoulos17}. The experiments by Pairam {\it et al.}~\cite{Pairam09} showed that torodial liquid droplets can break up into a precise number of droplets or only shrink towards its center to form a single spherical droplet, depending on the aspect ratio of the toroid. This demonstrates that the Rayleigh instability in the azimuthal direction and the shrinking instability in the radial direction are competing with each other to determine the dynamics of the toroid. This is a competition between the two time scales: one for toroid shrinkage towards its center and the other for neck pinch-off along the azimuthal direction. A toroid behaves like a cylinder when the aspect ratio (i.e., the ratio between the overall radius $R$ and the tube radius $a$) is large; this Rayleigh instability has been widely investigated. The shrinking instability, induced by the radial curvature, is a signature of the non-compact topology of the toroid and has not been well studied in literature, especially for solid-state dewetting.

\begin{figure}[!htp]
\centering
\includegraphics[width=0.48\textwidth]{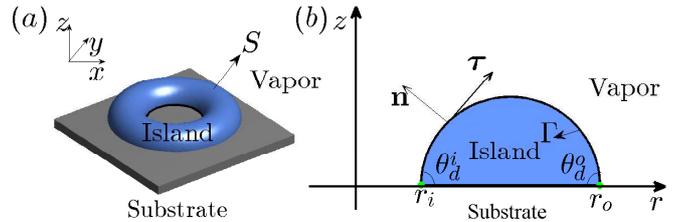}
\caption{(a) A schematic illustration of the solid-state dewetting of an initially, toroidal island on a flat, rigid substrate; (b) the cross-section profile (i.e., denoted as $\Gamma$) of the island is represented in a cylindrical coordinate system $(r,z)$, where $r_i,r_o$ representing the inner and outer contact points, respectively. Note that $\Gamma$ is not necessarily a circular arc during the evolution.}
\label{fig:det}
\end{figure}

In this paper, we apply Onsager's variational principle~\cite{Onsager31a,Onsager31b,Doi15} to derive a reduced-order model for analyzing the shrinking of a solid toroidal island on a solid substrate via surface diffusion.
To simplify the analysis, we explicitly assume that all interface/surface energies are isotropic, that elastic  (interface stress, stresses associated with capillarity) effects are negligible, and there are no chemical reactions or phase transformations occurring during the evolution.  We also focus on a pseudo-one dimensional morphological evolution example where transport occurs solely through surface diffusion (e.g., we neglect bulk diffusion and evaporation-condensation). None of these assumptions is essential and are employed in order to keep the presentation relatively simple, clear and analytically tractable. This application provides a concrete demonstration of the efficacy and simplicity of Onsager's variational principle for describing the surface diffusion-controlled morphology evolution problems, and applicability for analyzing even complex solid-state dewetting phenomena (e.g., a simultaneous consideration of the shrinking instability and Rayleigh instability).

This paper is organized as follows.
In Section 2, we present the set-up of the problem to be discussed here.
We also propose a full sharp-interface model for numerically simulating the problem.
In Section 3, we outline how to apply Onsager's variational principle to describe surface diffusion-controlled problems. In Section 4, we apply this variational principle to study the shrinkage of a solid toroid via surface diffusion. We derive an analytical expression for the shrinking rate of the toroidal island.
In Section 5, we present several comparisons between the predictions of the analytical formula and the numerical results from solving the full sharp-interface model. Finally, we draw some conclusions in Section 6.

\section{The geometry of the problem and its full sharp-interface model}

In this paper, we consider the shrinking dynamics of a solid toroidal island (in blue region) bonded to a flat rigid substrate  towards its center via a surface diffusion-controlled solid-state dewetting process, as illustrated in Fig.~\ref{fig:det}(a). We assume that the island maintains its axisymmetric shape during the evolution (i.e., we do not consider a Rayleigh instability along the azimuthal direction), and therefore, the film/vapor interface surface $S$ can be parameterized in cylindrical coordinate as (see Fig.~\ref{fig:det}(b))
\begin{equation}
S=(r(s,t)\cos\varphi,~r(s,t)\sin\varphi,~z(s,t)),
\end{equation}
where $r(s,t)$ is the radial distance, $z(s,t)$ is the local height with $s$ representing the arc length of the cross-section profile $\Gamma=(r(s,t),z(s,t))$ of the surface $S$, and $\varphi\in[0,2\pi]$ is the azimuthal angle.

The total interfacial free energy of the system can be described as (up to a constant)~\cite{Bao17b,Jiang18,Zhao18}
\begin{equation}
    W = \iint\limits_S \gamma_{_{\subFV}} \,dS +\underbrace{(\gamma_{_{\subFS}} - \gamma_{_{\subVS}})\pi(r_o^2-r_i^2)}_{\rm{Substrate}\,\,\rm{Energy}},
    \label{eqn:energy}
\end{equation}
where the constants $\gamma_{_{\subFV}}$ (i.e., $\gamma_0$), $\gamma_{_{\subFS}}$ and $\gamma_{_{\subVS}}$ represent the film/vapor, film/substrate and vapor/substrate surface energy densities, respectively, and $r_o$ and $r_i$ are the radii of the outer and inner contact lines, respectively (shown in Fig.~\ref{fig:det}(b)).

Because we assume that the surface is axisymmetric, the island morphology evolution can be described in terms of the evolution of the cross-section curve $\Gamma$.
For brevity, we denote $\Gamma(t)=\vec X(s,t)=(r(s,t),z(s,t))$ as the cross-section profile of the surface $S$ with $0\leq s\leq L(t)$ and $r(0,t)=r_i,r(L,t)=r_o$.
Based on a thermodynamic variational analysis, we previously  proposed a sharp-interface model for simulating solid-state dewetting in three dimensions for axisymmetric geometries~\cite{Zhao18}.
By choosing a length scale and surface energy density scale for normalization as $L_0$ and $\gamma_{_0}$ respectively,  the time normalized by ${L^4_0}/{(B\gamma_{_0})}$, and the contact line mobility by ${B}/{L^3_0}$,
this leads to the following dimensionless sharp-interface evolution model (for isotropic surface energy)~\cite{Zhao18}:
\begin{eqnarray}
&&\partial_{\tilde{t}}\tilde{\vec {X}}=\frac{1}{\tilde{r}}\partial_{\tilde{s}}\big(\tilde{r}\partial_{\tilde{s}}\tilde{\mu}\big)\,\vec{n},\quad 0 < \tilde{s} < \tilde{L}(\tilde{t}),\,\tilde{t}>0,\\
&&\tilde{\mu}=\tilde{\mathcal{H}}=\tilde{\kappa}-\frac{\partial_{\tilde{s}} \tilde{z}}{\tilde{r}},\quad \tilde{\kappa}=-(\partial_{\tilde{s}\tilde{s}}\tilde{\vec X})\cdot\vec n;
\end{eqnarray}
where $\tilde{\mu}$ is the (dimensionless) chemical potential, $\tilde{\mathcal{H}}$ is the mean curvature of the surface $S$, and $\tilde{\kappa}$ is the curvature of the curve $\Gamma$, $\vec n=(-\partial_{\tilde{s}} \tilde{z},\partial_{\tilde{s}} \tilde{r})$ is the outer unit normal vector of the curve $\Gamma$. Note that we refer to dimensionless physical quantities with tildes in order to distinguish from their dimensional counterparts.

The above equations are subject to the following dimensionless boundary conditions:
\begin{itemize}
\item[(i)] Contact line condition
\begin{equation}
\tilde{z}(0,\tilde{t}) = 0,\quad \tilde{z}(\tilde{L}, \tilde{t}) = 0;
\label{eqn:pdebc1}
\end{equation}
\item[(ii)] Relaxed contact angle condition
\begin{eqnarray}
\label{eqn:pdebc2}
&&\frac{d \tilde{r}_i}{d \tilde{t}}= \tilde{\eta}(\cos\theta_d^i-\cos\theta_i),\\
&&\frac{d \tilde{r}_o}{d \tilde{t}}= -\tilde{\eta}(\cos\theta_d^o-\cos\theta_i);
\end{eqnarray}
\item[(iii)] Zero-mass flux condition
\begin{equation}
\partial_{\tilde{s}} \tilde{\mu}(0, \tilde{t})=0,\quad  \partial_{\tilde{s}}\tilde{\mu}(\tilde{L}, \tilde{t}) = 0.
\label{eqn:pdebc3}
\end{equation}
\end{itemize}
Here $\theta_d^i\,,\theta_d^o$ are the (dynamic) contact angles for the inner and outer contact lines,
respectively, $0<\tilde{\eta}<+\infty$ denotes the (dimensionless) contact line mobility~\cite{Wang15,Upmanyu02}, and $\theta_i$ is the isotropic equilibrium (Young) contact angle, i.e., $\cos\theta_i=(\gamma_{_{\subVS}}-\gamma_{_{\subFS}})/ \gamma_0$. Boundary condition (i) ensures that contact lines always move along the substrate, (ii) describes the dynamic relaxation of the contact angle, and (iii) ensures that the total volume/mass of the island is conserved (i.e., no mass flux into/from the island at the contact lines)~\cite{Wang15,Jiang16}.

Note that when the contact line mobility $\tilde{\eta}$ goes to infinity, $\theta_d^i=\theta_d^o=\theta_i$ (since the velocity of moving contact lines is finite). In this limit, boundary condition (ii) reduces to a fixed contact angle condition and  contact line motion will not dissipate any free energy. We use the relaxed contact angle boundary condition (ii) in our numerical computations because it can improve numerical stability and, in the fast contact line motion limit, it is also physically important. The accurate, efficient parametric finite element method for numerically solving the above sharp-interface model is described in~\cite{Bao17,Zhao18}.

In general, it would be impossible to directly obtain analytical solutions for the above full sharp-interface model.
In the following sections, we develop an Onsager's variational principle approach to derive a reduced-order model for describing the dynamics of a toroidal island shrinking via surface diffusion. We validate the resultant analytical approach by comparison with the numerical results from solving the above full sharp-interface model.

\section{Onsager's variational principle}

Onsager's variational principle, first formulated by Lars Onsager in 1931~\cite{Onsager31a,Onsager31b}, is based on the reciprocal relations in linear irreversible thermodynamics.
This variational principle has been found to be very useful in deriving  evolution equations in fluid dynamics~\cite{Qian06,Qian17,Xu16,Di18,Man16} and soft matter physics~\cite{Doi11,Doi13book,Doi15}.
In this paper, we adopt this variational approach to solid-state dewetting problems in materials science.
In particular, we apply it to the case where the evolution is controlled by  surface diffusion.

Consider an isothermal system which may include various boundaries (boundaries between the solid island and solid substrate, the solid phases and the vapor).
If the system is away from its equilibrium state, then spontaneous processes will occur that tend to establish equilibrium. In the linear response regime, the evolution of the system is governed by the following principle.
Let $\alpha(t)=(\alpha_1(t),\alpha_2(t),\ldots,\alpha_n(t))$ be a set of state variables.
The time evolution of the system, given by the time derivatives of these state variables
$\dot{\alpha}(t)=(\dot{\alpha}_1(t),\dot{\alpha}_2(t),\ldots,\dot{\alpha}_n(t))$, is determined by minimizing the following (Rayleighian) action with respect to the rates $\{\dot{\alpha}_i\}$~\cite{Doi15,Xu16,Suo97}:
\begin{equation}
\mathcal{R}(\dot{\alpha},\alpha)=\dot{W}(\alpha, \dot{\alpha})+\Phi(\dot{\alpha},\dot{\alpha}).
\label{eqn:rayleigh}
\end{equation}
Here $W(\alpha):= W(\alpha_1,\alpha_2,\ldots,\alpha_n)$ represents the total free energy of the system,
$\dot{W}$ is the rate of change of the total free energy
\begin{equation}
\dot{W}(\alpha, \dot{\alpha})=\sum_{i}\frac{\partial W}{\partial \alpha_i}\dot{\alpha_i},
\end{equation}
where the symbol ``$\cdot$'' denotes time derivatives, and $\Phi(\dot{\alpha},\dot{\alpha})$ is the free energy dissipation function which is defined as half the rate of free energy dissipation.
In the linear response regime, the dissipation function can be written as a quadratic function of the rates $\{\dot{\alpha}_i\}$, i.e.,
\begin{equation}
\Phi(\dot{\alpha},\dot{\alpha})=\frac{1}{2}\sum_{i,j}\zeta_{ij}(\alpha)\dot{\alpha}_i\dot{\alpha}_j,
\end{equation}
with the friction coefficients $\zeta_{ij}$ forming a symmetric and positive definite matrix.
Minimizing the Rayleighian in Eq.~\eqref{eqn:rayleigh} with respect to the rates $\{\dot{\alpha}_i\}$ yields the kinetic equations:
\begin{equation}
-\frac{\partial W}{\partial \alpha_i}=\sum_{j}\zeta_{ij}\dot{\alpha}_j,\qquad i=1,2,\ldots,n.
\end{equation}
This describes the force balance between the reversible force $-\frac{\partial W}{\partial \alpha_i}$ and the dissipative force $\frac{\partial \Phi}{\partial \dot{\alpha}_i}$ which is linear in the rates $\{\dot{\alpha}_i\}$.
A simple calculation shows that the variational principle leads to $\dot{W}=-2\Phi$, which means $\Phi$ is half the rate of free energy dissipation. Physically, the variational principle outlined above for {\it isothermal} systems can be derived from the maximization of the Onsager-Machlup action with respect to the rates $\{\dot{\alpha}_i\}$ for more general {\it non-isothermal} systems~\cite{Onsager31a,Onsager31b}.

Now consider a dissipative system that can be (approximately) described by a finite set of state variables.
If we compute its total free energy $W$ and dissipation function $\Phi$ as a function of these state variables
and their time derivatives, then application of Onsager's variational principle yields a system of {\it ordinary differential equations} (ODEs) that describe the time evolution of the state variables, i.e., the time evolution of the system~\cite{Xu16,Man16,Doi15}.
The purpose of the present work is to apply Onsager's variational principle to solid-state dewetting (in particular, we focus on axisymmetric toroidal islands on a substrate).
We first derive a reduced-order variational model to describe the time evolution of the island morphology.
We then compare the analytical results with numerical results obtained by the solutions of the full sharp-interface model.

We note that the full surface diffusion equation can also be directly derived from Onsager's variational principle.
This makes a reduced variational model possible and viable for surface diffusion controlled problems.
In his pioneering work for surface diffusion-controlled morphology evolution problems~\cite{Mullins57}, Mullins made use of three important relations:
\begin{eqnarray}
\mu&=&\Omega_0\gamma_0\mathcal{H}, \label{eqn:mulone}\\
\vec{V}&=&-\frac{D_s}{k_BT}\nabla_s\mu,\label{eqn:multwo}\\
v_n&=&-\Omega_0\nabla_s\cdot(\nu\vec{V}). \label{eqn:multhree}
\end{eqnarray}
The surface diffusion equation, i.e., Eq.~\eqref{eqn:mullins}, is obtained by combining these relations.
Here (i) $\mu$ is the chemical potential defined by Eq.~\eqref{eqn:mulone} according to the variation of the total free energy $W=\iint_S \gamma_0\,dS$, (ii) $\vec{V}$ is the velocity of atoms along the surface driven by the gradient of chemical potential according to Eq.~\eqref{eqn:multwo} (this is a constitutive equation that can be derived from Onsager's principle), and (iii) $v_n$ is the normal velocity of the interface and $\nu$ is the number of diffusing atoms per unit area (Eq.~\eqref{eqn:multhree} is simply a continuity equation for mass conservation during surface diffusion).
Note that $-\nabla_s \cdot (\nu \vec{V})$ represents the rate of change of the number of particles per unit area, $-\Omega_0 \nabla_s \cdot (\nu \vec{V})$ is the rate of change of the volume of particles per unit area (i.e., the rate of change of the height normal to the film-vapor interface).

In order to derive Eq.~\eqref{eqn:multwo}, the rate of change of the total free energy can be calculated as
\begin{eqnarray*}
\dot{W}&=&\iint\nolimits_S \gamma_0\,\mathcal{H}\,v_n\,dS \\
&=&\iint\nolimits_S \gamma_0\,\mathcal{H}\,[-\Omega_0 \nabla_s \cdot (\nu \vec{V})]\,dS \\
&=&\iint\nolimits_S \mu \, [-\nabla_s \cdot (\nu \vec{V})]\,dS \\
&=&\iint\nolimits_S (\nabla_s \mu) \cdot (\nu \vec{V})\,dS,
\end{eqnarray*}
where an integration by parts has been used. For surface diffusion,
the dissipation function is defined as
\begin{equation*}
\Phi = \frac{1}{2}\iint\nolimits_S  \frac{k_B T}{D_s} |\vec {V}|^2 \nu\,dS,
\end{equation*}
where ${k_B T}/{D_s}$ is the friction coefficient for diffusing atoms.
Defining a Rayleighian $\mathcal{R}=\dot{W}+\Phi$ and minimizing $\mathcal{R}$ with respect to the velocity $\vec{V}$, we obtain the constitutive equation for surface diffusion, i.e., Eq.~\eqref{eqn:multwo}.

\section{A reduced-order variational model}

\begin{figure}[!htp]
\centering
\includegraphics[width=0.48\textwidth]{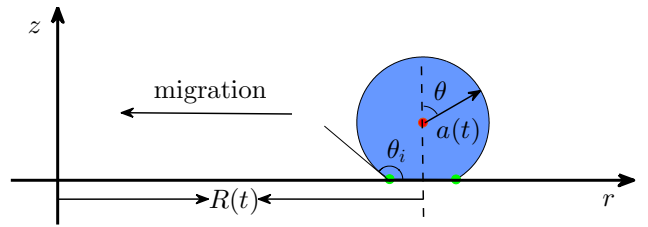}
\caption{A cross-section profile of solid-state dewetting of an initially, toroidal island on a flat, rigid substrate in cylindrical coordinates, where $a:=a(t)$ is the radius of the circle (or toroid), $R:=R(t)$ is the overall radius (i.e., the distance between the origin of the circle and the $z$-axis).  We assume that the ratio $a(t)/R(t)$ is not large.}
\label{fig:det2}
\end{figure}

In the reduced-order variational model, we assume that (i) during the island evolution (see Fig.~\ref{fig:det2}), the cross-section profile of the island is a circular arc which meets the substrate at the isotropic Young contact angle $\theta_i \in [0,\pi]$  and (ii) the contact line does not dissipate any free energy when it moves along the substrate. As noted above,  we do not consider the instability along the azimuthal direction (i.e., Rayleigh instability) here.

We express the cross-section profile of the island $\Gamma(t):=(r(\theta,t), z(\theta,t))$ in cylindrical coordinates $(r,z)$ as
\begin{equation}
\left\{
\begin{array}{l}
r(\theta,t)=R(t)+a(t)\sin\theta, \\[0.8em]
z(\theta,t)=a(t)(\cos\theta-\cos\theta_i),
\end{array}
\right.
\theta\in[-\theta_i,\theta_i],
\label{eqn:profile}
\end{equation}
where $\theta$ is a parametrization of the cross-section curve $\Gamma$.
The three-dimensional surface profile of the island is obtained by rotating $\Gamma$ around the $z$-axis, i.e., $S(t)=(r(\theta,t)\cos\varphi,~r(\theta,t)\sin\varphi,~z(\theta,t))$, where $\varphi \in [0,2\pi]$ is the azimuthal angle. Since the total volume of the solid island is conserved during the evolution, the initial volume of the island  $V_0$, i.e.,
\begin{equation}
V_0=\pi R a^2(2\theta_i-\sin 2\theta_i),
\label{eqn:volume}
\end{equation}
constrains the state variables $a:=a(t)$ and $R:=R(t)$.
Use of this constraint implies that there is only one independent parameter, i.e., $R$ or $a$.
Without loss of generality, we consider the evolution of the island in terms of the parameter $R$  and define the Rayleighian  of the system $\mathcal{R}$ as
\begin{equation}
\mathcal{R}= \dot{W}(R, \dot{R}) + \Phi(\dot{R}, \dot{R}).
\label{eqn:Onsager}
\end{equation}
Minimizing the Rayleighian with respect to the rate $\dot{R}$  yields the time evolution equation of $R$.
We now determine the total free energy function $W$ and the dissipation function $\Phi$ of the system, respectively.

We write the total interfacial free energy of the system in terms of the state variable $R$, using Eq.~\eqref{eqn:energy}, as
\begin{equation}
W=\gamma_0\Bigl[2\pi R\,a\bigl(2\theta_i-\sin 2\theta_i \bigr)\Bigr].
\label{eqn:Massconservation}
\end{equation}
By using Eq.~\eqref{eqn:volume}, we obtain $a=\sqrt{{V_0}/{\pi R(2\theta_i-\sin 2\theta_i)}}$.
Substituting this relation into Eq.~\eqref{eqn:Massconservation}, and taking the time derivative, we  obtain
\begin{equation}
\dot{W}=\gamma_0\sqrt{\pi V_0 (2\theta_i-\sin 2\theta_i)}\,R^{-\frac{1}{2}}\dot{R}.
\label{eqn:derivative1}
\end{equation}

The island morphology evolution is driven by the interfacial free energy minimization and its mass transport occurs through surface-diffusion. The resulting dissipation function can be written as
\begin{eqnarray}
\Phi &=& \frac{1}{2}\iint\nolimits_S \frac{k_B T}{D_s} |\vec V|^2 \nu\;dS \nonumber \\[0.5em]
     &=& \frac{1}{2}\frac{k_B T}{D_s\nu}\iint\nolimits_S  |\vec J|^2 \;dS.
\label{eqn:PhiJ}
\end{eqnarray}
where $\vec J$ is the surface current (or mass flux) of atoms along the interface~\cite{Mullins57},
\begin{equation}
\vec J=\nu \vec V=-\frac{D_s\nu}{k_B T}\nabla_s\mu,
\label{eqn:J}
\end{equation}
and $\nabla_s$ is the surface gradient operator.
By multiplying by the atomic volume $\Omega_0$, the mass flux can be converted to the normal velocity $v_n$ of the surface element,
\begin{equation}
v_n=-\Omega_0\,(\nabla_s \cdot \vec{J}).
\label{eqn:velo}
\end{equation}

Because the island shape is assumed to be axisymmetric, the normal velocity $v_n$ of the island can be calculated by restricting the problem to the cross-section profile $\Gamma$ (shown in Fig.~\ref{fig:det2}).
By taking the time derivative of the surface profile $\Gamma$, i.e., Eq.~\eqref{eqn:profile} and
multiplying by the surface unit normal vector $\vec n=(\sin\theta,\cos\theta)$, we  obtain
the normal velocity $v_n$ of the surface element along the cross-section curve $\Gamma$:
\begin{equation}
v_n(\theta)=\dot{R}\sin\theta+\dot{a}(1-\cos\theta_i\cos\theta), \quad \theta \in [-\theta_i, \theta_i].
\label{eqn:surfacevelocity}
\end{equation}
The mass flux vector $\vec J$ is only parallel to the tangential to $\Gamma$; hence, $\vec J=J\boldsymbol{\tau}$ where $\boldsymbol{\tau}$ representing the unit tangential vector, where $J$ is the magnitude of the mass flux vector.
The surface divergence of the flux is related to the surface velocity by
\begin{equation}
-\Omega_0 \big[\nabla_s\cdot (J\boldsymbol{\tau})\big]=-\Omega_0 \frac{\partial_s(rJ)}{r} = v_n(\theta),
\label{eqn:ODEflux}
\end{equation}
where $\partial_s$ represents the first-order derivative with respect to the arc length of $\Gamma$.
By integrating both sides and applying the zero-mass flux boundary condition $J(-\theta_i)=0$, the magnitude of the mass-flux is
\begin{equation}
J(\theta)=-\frac{1}{r\Omega_0}\int_{-\theta_i}^\theta r\,v_n(\theta)\, a\,d\theta.
\label{eqn:fluxadd}
\end{equation}

Inserting the expression $v_n$ from Eq.~\eqref{eqn:surfacevelocity} and the expression $r=R+a\sin\theta$ into the above Eq.\eqref{eqn:fluxadd}, the mass flux magnitude can be reformulated as:
\begin{align}
J(\theta)&=-\frac{a}{\Omega_0(R+a\sin\theta)}\Big[R\dot{R}\int_{-\theta_i}^\theta\sin\theta d\theta \nonumber\\
&+a\dot{R}\int_{-\theta_i}^\theta\sin^2\theta\;d\theta + R\dot{a}\int_{-\theta_i}^\theta(1-\cos\theta_i\cos\theta)d\theta \nonumber\\
&+ a\dot{a}\int_{-\theta_i}^\theta \sin\theta(1-\cos\theta_i\cos\theta)\;d\theta\Big].
\label{eqn:longform}
\end{align}
Taking the time derivative of Eq.~\eqref{eqn:volume} yields
\begin{equation}
a\dot{R}=-2R\dot{a}.
\label{eqn:Randa}
\end{equation}
Since $\delta=\frac{a}{R}\ll1$, and making use of Eq.~\eqref{eqn:Randa}, we can reformulate the mass flux magnitude $J(\theta)$ in terms of $\delta$ as the following form,
\begin{equation}
J(\theta)=\frac{\cos\theta-\cos\theta_i}{\Omega_0}\Big[1-\frac{1}{2}\delta\sin\theta+\mathcal{O}(\delta^2)\Big]a\dot{R}.
\end{equation}
Substituting this expression for $J(\theta)$ into Eq.~\eqref{eqn:PhiJ}, we obtain the dissipation function to leading-order as
\begin{align}
\Phi &=\frac{1}{2}\frac{k_B T}{D_s\nu}\iint\nolimits_S |\vec J|^2\;dS \nonumber \\
&= \frac{1}{2}\frac{k_B T}{D_s\nu}\int\nolimits_0^{2\pi}\,r\,d\varphi\,\int\nolimits_{-\theta_i}^{\theta_i} J^2(\theta)\,a\,d\theta \nonumber \\
&= \frac{k_B T}{D_s\nu}\pi a\int_{-\theta_i}^{\theta_i}(R+a\sin\theta) J^2(\theta) \,d\theta \nonumber\\
&= \frac{k_B T}{D_s\nu}\pi Ra\int_{-\theta_i}^{\theta_i}(1+\delta\sin\theta)J^2(\theta) \,d\theta, \nonumber \\
&= \frac{k_B T}{D_s\nu\Omega_0^2}\pi Ra^3\dot{R}^2\Bigl[g(\theta_i)+\mathcal{O}(\delta^2)\Bigr],
\label{eqn:derivative2}
\end{align}
where
\begin{equation}
g(\theta_i)=\theta_i(2+\cos 2\theta_i)-\frac{3}{2}\sin 2\theta_i.
\end{equation}

With the time derivative of the total free energy function $\dot{W}$ (Eq.~\eqref{eqn:derivative1})
and the dissipation function $\Phi$ (Eq.~\eqref{eqn:derivative2}) in terms of $R$ and $\dot{R}$, we apply Onsager's variational principle by minimizing the Rayleighian function $\mathcal{R}$ (Eq.~\eqref{eqn:Onsager}) with respect to $\dot{R}$. The resultant kinetic equation for the shrinking rate $v$ is found, to leading order, to be
\begin{equation}
v=-\dot{R}(t)\approx C(\theta_i)\,\frac{B\gamma_0}{V_0},
\label{eqn:v}
\end{equation}
where
\begin{equation}
C(\theta_i)=\frac{\pi(2\theta_i-\sin 2\theta_i)^2}{2\,g(\theta_i)}.
\label{eqn:ctheta}
\end{equation}
The torus shrinking rate is (to leading order) proportional to the material constants $B$ and $\gamma_0$, and inversely proportional to the volume of the toroid $V_0$, and the coefficient $C(\theta_i)$  only depends on the isotropic Young angle $\theta_i$.
Given the initial toroidal island location $R_0$, we find that the toroidal island evolves as
\begin{equation}
R(t)=R_0-C(\theta_i)\,\frac{B\gamma_0}{V_0}t,\qquad 0\leq t < t_\delta(\theta_i),
\label{eqn:R(t)}
\end{equation}
where $t_\delta(\theta_i)$ represents a time where the leading-order approximation breaks down (i.e., where the assumption that $\delta={a}/{R}\ll1$ breaks down), and this time depends on $\theta_i$. As an aside, we note that the constant shrinking speed found here is similar to experimental observations on the shrinking speed of toroidal liquid droplets~\cite{Pairam09}.

\section{Comparison with numerical results}

In order to validate the variational model, we compare our predictions with the results of numerical simulations based on the full sharp-interface model. In the following numerical simulations, we choose a large contact line mobility (e.g., $\tilde{\eta}=100$) to ensure that the contact angle is always near its equilibrium value and the dissipation associated with contact line motion is negligible small as possible, and the initial shape of the toroidal island is chosen as half of a torus (i.e., the initial contact angle is $\pi/2$).

\begin{figure}[!htp]
\centering
\includegraphics[width=0.46\textwidth]{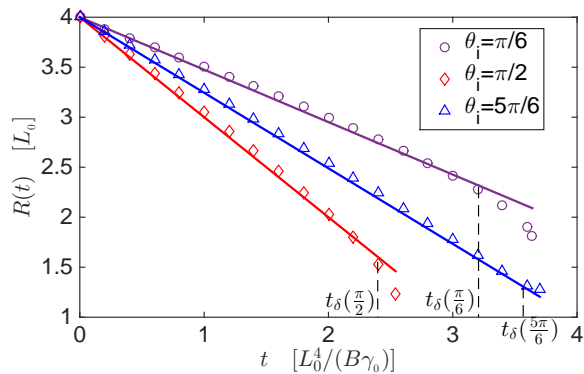}
\caption{Comparisons between the numerical results of $R(t)$ by solving the full sharp-interface model as described in Section 2 and the variational prediction (Eq.~\eqref{eqn:R(t)}). The ``circles'', ``rhombi'' and ``triangles''  are  numerical results obtained from solving the full model and the solid lines are the predicted formula for different isotropic Young angles. The initial parameters are chosen as $R_0=4.0$, $a_0=0.5$, and $L_0$ is the length scale.}
\label{fig:det3}
\end{figure}

Our numerical simulations under the conditions represented in Fig.~\ref{fig:det3} demonstrate that assumption (i), that the cross-section profile $\Gamma(t)$ is always a circular arc, is valid.
Taking $\theta_i=\pi/2$ as an example, we find that this assumption is valid for about $\delta \leq 1/3$.
We also found that this assumption is valid up to the time when $R(t)$ begins to deviate from the predicted trajectory by the analytical formula in Eq.~\eqref{eqn:R(t)} (see Fig.~\ref{fig:det3}).

Figure~\ref{fig:det3} shows a comparison of  $R(t)$ from the numerical simulation results based upon the full sharp-interface model as well as the analytical results from the variational model Eq.~\eqref{eqn:R(t)} for three different isotropic Young angles $\theta_i$.
As shown in the figure,  the numerical results (symbols) for $R(t)$ are in very good agreement with the analytical predictions (solid lines) for all contact angles from the beginning to the late time $t_\delta(\theta_i)$.
Our numerical simulation also indicate that when $\delta=a(t)/R(t)$ is small (not shown), the toroid shrinks towards its center in a quasi-static manner, i.e., its cross-section profile remains nearly a circular arc, consistent with our assumption in the analysis.
However, as time evolves, $\delta$ increases to a value at which the cross-section profile begins to show non-negligible deviations from the circular arc assumption and the simple analytical result Eq.~\eqref{eqn:R(t)} breaks down.
\begin{figure}[!htp]
\centering
\includegraphics[width=0.46\textwidth]{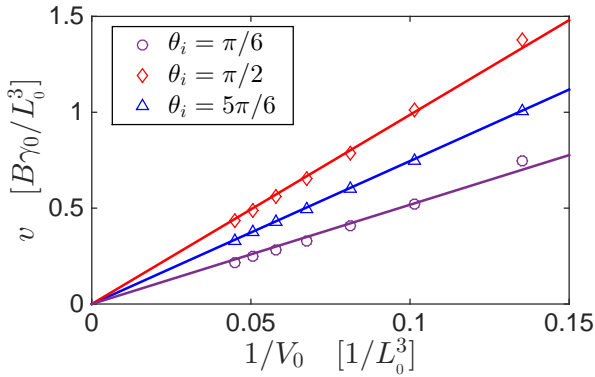}
\caption{Comparisons between the rate at which the toroidal island shrinks $v$ obtained from solving the full sharp-interface model (symbols) and the theoretical shrinking speed, by Eq.~\eqref{eqn:v} (solid lines) as a function of the island volume $V_0$ for different isotropic Young angles $\theta_i$.}
\label{fig:det4}
\end{figure}

We performed a least-square linear fitting to numerical data of $R(t)$ obtained from solving the full sharp-interface model for $0\leq t \leq t_\delta(\theta_i)$ in order to determine the variation of toroidal island shrinking speeds for several isotropic Young angles $\theta_i$ and initial volumes $V_0$.
Figure~\ref{fig:det4} shows the comparisons between the rate of island shrinking from the numerical simulations and analytical predictions, where the values of $C(\theta_i)$ were estimated from the numerical results.
This figure shows that the numerical results suggest that $v$ is inversely proportional to the toroidal island volume $V_0$ as predicted analytically by Eq.~\eqref{eqn:v}.

\begin{figure}[!htp]
\centering
\includegraphics[width=0.48\textwidth]{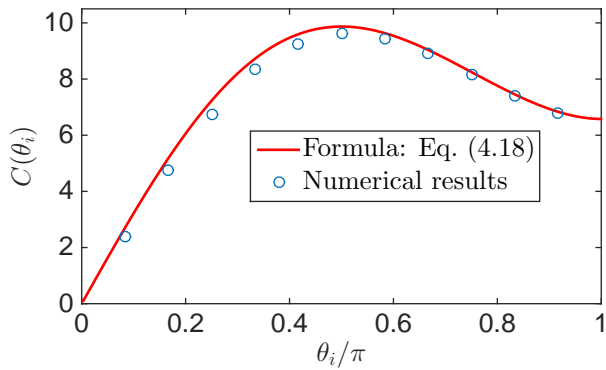}
\caption{Comparisons between the fitting values of $C(\theta_i)$ (shown in ``circles'') obtained by numerically solving the full sharp-interface model and the analytical values given by Eq.~\eqref{eqn:ctheta} (shown in red solid line).}
\label{fig:det5}
\end{figure}

Figure~\ref{fig:det5} shows some comparisons between $C(\theta_i)$ obtained from numerical results (shown in ``circles'') and the analytical expression in Eq.~\eqref{eqn:ctheta}.
Again, we see that the numerical results based upon the full model are accurately predicted by the analytical results based upon the variational model.
These results show that the function $C(\theta_i)$ reaches the maximum value when $\theta_i=\pi/2$; i.e., the toroidal island shrinks fastest when $\theta_i=\pi/2$.

\section{Conclusions}

We considered the evolution of a solid toroidal island on a flat substrate, evolving by capillarity-driven surface diffusion. This problem is an example of the evolution of the complex local geometric features often observed in dewetting of a solid film on a substrate.
Our general approach to such problems as that described here, is based upon Onsager's variational principle.
Using this approach, we derived a reduced-order variational model for describing and analyzing the shrinking of a toroidal island. We obtained an analytical formula for the rate at which the island shrinks; the shrinking rate is proportional to the material constants $B=\frac{D_s \nu\,\Omega_0^2}{k_BT}$ and $\gamma_0$, and inversely proportional to the island volume $V_0$. The analytical predictions are validated by detailed comparisons with accurate numerical simulations based upon a full sharp-interface model; and the agreement is excellent.

To our knowledge, the present work is the first demonstration of Onsager's variational principle for describing surface diffusion-controlled problems in materials science. We plan to use this principle to investigate more complicated phenomena, including analyzing Rayleigh instabilities of islands on substrates~\cite{Thompson12,Kim15}, the migration of ``small'' particles on curved substrates~\cite{Jiang18b}, and power-law scaling of a retracting semi-infinite thin film~\cite{Zucker16b}.

\section*{Acknowledgements}

This work was partially supported by the National Natural Science Foundation of China Nos.~11871384 (W.J.),
91630313 (W.J.) and 91630207 (W.B.), Natural Science Foundation of Hubei Province No.~2018CFB466 (W.J.),
Hong Kong RGC CRF grant No.~C1018-17G (T.Q.), the Academic Research Fund of the Ministry of
Education of Singapore grant No.~R-146-000-247-114 (Q.Z.\& W.B.) and the NSF Division of Materials Research through Award 1507013 (D.J.S.). The first author wishes to thank Professor Xian-min Xu for helpful discussions.
This work was partially done while
the authors were visiting the Institute for Mathematical
Sciences, National University of Singapore, in 2018.

\section*{References}

\end{document}